\documentclass[runningheads]{llncs}

\usepackage{amsmath}
\usepackage{graphicx}
\usepackage{booktabs}
\usepackage{xcolor}
\usepackage{rotating}
\usepackage[hidelinks]{hyperref}
\usepackage{xurl}
\usepackage{microtype}
\usepackage{newunicodechar}
\usepackage{textcomp}
\newunicodechar{Δ}{\ensuremath{\Delta}}
\newunicodechar{α}{\ensuremath{\alpha}}
\newunicodechar{β}{\ensuremath{\beta}}
\newunicodechar{ε}{\ensuremath{\varepsilon}}
\newunicodechar{δ}{\ensuremath{\delta}}
\newunicodechar{√}{\ensuremath{\surd}}
\newunicodechar{×}{\ensuremath{\times}}
\newunicodechar{≥}{\ensuremath{\geq}}
\newunicodechar{≤}{\ensuremath{\leq}}
\newunicodechar{≈}{\ensuremath{\approx}}
\newunicodechar{∈}{\ensuremath{\in}}
\newunicodechar{→}{\ensuremath{\rightarrow}}
\newunicodechar{·}{\ensuremath{\cdot}}
\newunicodechar{−}{\ensuremath{-}}

\begin{document}

\title{Static PTX Metrics Track Structural Kernel\\Regressions but Miss Semantic Ones}
\titlerunning{Static PTX Metrics for Kernel Regressions}

\author{Dipankar Sarkar\orcidID{0000-0001-5431-6367}}
\authorrunning{D. Sarkar}
\institute{Arizona State University, USA \\
\email{dsarkar3@asu.edu}}

\maketitle

\begin{abstract}
We pair each GPU kernel's static PTX metrics (registers, spills,
instruction count) with CUDA-event-timed runtime on five GPU classes:
RTX~3060, A10, L40S, A100~SXM4, and H100~NVL. In this corpus and
toolchain the static and measured signals separate cleanly along one
axis. Per-pair $\Delta$regs and $\Delta$instrs are identical across
all five GPUs for any given (correct, buggy) pair. Measured
$\Delta$perf\% is not. Structural bugs that change the kernel's work
are unambiguous in the static signal. The
\texttt{gelu\_triton\_buggy} variant, which drops a leading $0.5$
factor, removes 8 instructions and 8 registers. The corresponding
measured $\Delta$perf\% on RTX~3060 is $+3.2\%$, within the
run-to-run noise band at the sub-millisecond scale these corpus
kernels occupy. Semantic bugs that swap one constant for another are
invisible to the static signal. The \texttt{softmax\_triton\_buggy}
variant, which substitutes \texttt{other=0.0} for \texttt{-inf} on
the masked load, compiles to byte-identical PTX. The paper's bounded
claim is that, for this corpus and toolchain, a static-PTX delta
gate is a portable pre-filter that separates structural from
semantic changes; measured runtime deltas at this scale are
hardware- and noise-sensitive and are not a substitute.
\end{abstract}

\section{Introduction}
\label{sec:intro}

CI gating on GPU kernel changes is expensive. Running every variant
on real hardware costs seconds to minutes per test. A long-standing
folk view is that static PTX metrics (register count, spills to local
memory, instruction count) are leading indicators of measured GPU
performance and can therefore gate CI without hardware execution. We
test the claim on a controlled corpus.

The claim is a bounded one. Static PTX metrics track structural
changes to the kernel's compiled work envelope. In this corpus,
structural bugs produce nonzero $\Delta$regs or $\Delta$instrs,
while semantic-only constant changes produce zero static delta.
Measured runtime deltas are a different signal: at the
sub-millisecond scale of these kernels, CUDA-event timing is
dominated by launch and host variance, so measured $\Delta$perf\%
does not reliably separate the two bug classes without larger
shapes or per-architecture calibration. We give a measured example
of each. The cross-architecture sweep in
Section~\ref{sec:eval:cross} strengthens the portability side of
the static claim. The static signal is identical across five GPU
classes for the same kernel because it is determined at compile
time.

\section{Related Work}
\label{sec:related}

\textbf{GPU performance modeling.} The literature on GPU performance
prediction from compiler-level information is rich
\cite{gpgpu_latency2019,rndn2023}. The dominant approach uses a
microarchitectural model (occupancy, memory bandwidth, instruction mix)
to predict runtime. Correlation with measured performance is typically
used to validate the model.

\textbf{Register pressure and spills.} When register pressure exceeds
the SM's per-thread limit, ptxas spills to local memory backed by
global DRAM \cite{ptxasforum,regdem2019}. This causes
correctness-preserving but performance-destroying load and store
traffic. Static spill detection through \texttt{ld.local} and
\texttt{st.local} patterns in PTX is the typical CI gate.

\textbf{Triton-level optimisation.} Triton~\cite{triton2019} compiles
Python-level kernel descriptions to PTX with autotuning over
\texttt{BLOCK\_M}, \texttt{BLOCK\_N}, and \texttt{num\_warps}.
Library-level wrappers further trade off register pressure against
tiling.

\textbf{The gap.} No prior work, to our knowledge, controls for kernel
semantics while varying static PTX metrics, and measures the
regression-prediction correlation as a function of bug class. We do.

\section{Method}
\label{sec:method}

\subsection{Static metrics}
\label{sec:method:static}

\texttt{crates/gpuemu-daemon/src/artifact.rs} parses PTX text and
reports five metrics.

\begin{itemize}
\item \texttt{register\_count}: total number of declared registers
across all bank types.
\item \texttt{spill\_count}: count of \texttt{ld.local} and
\texttt{st.local} mnemonics, used as a proxy for spill traffic.
\item \texttt{local\_memory\_bytes}: declared \texttt{.local}
allocation sizes.
\item \texttt{instruction\_count}: count of indented PTX source
lines whose first non-whitespace token is a lowercase mnemonic with
optional \texttt{.}-modifiers (e.g.\ \texttt{add.f32},
\texttt{mov.b32}). Labels, \texttt{.}-prefixed directives,
comments, \texttt{.entry}/\texttt{.func} headers, and
predicate-prefixed lines (\texttt{@\%p1 ...}) do not contribute.
The same counter is applied uniformly across all kernel pairs.
\item \texttt{patterns\_found}: the set of instruction mnemonics for
required and forbidden-pattern policies.
\end{itemize}

The analyser is invoked through the daemon's \texttt{LintKernel} RPC.
The same artifact-analyser code is shared across all gpuemu use cases.

\subsection{Measured perf}
\label{sec:method:perf}

\texttt{drivers/\_capture.py} wraps a kernel launch with CUDA-event
timing and a fixed warmup and iteration count. It reports
\texttt{ms\_min}, \texttt{ms\_median}, and \texttt{ms\_mean}. Device
identity (name, SM count, capability) is also recorded.

\subsection{The pairing protocol}
\label{sec:method:protocol}

For each Triton kernel in the corpus, the P4 driver
(\texttt{drivers/p4\_artifacts.py}) performs five steps.

\begin{enumerate}
\item Clear the Triton cache. The produced PTX is then attributable to
this kernel only.
\item Run \texttt{kernel.run(inputs)} on a representative shape to
populate the cache.
\item Send each emitted \texttt{.ptx} to the daemon's
\texttt{lint\_kernel} RPC for static metrics.
\item Wrap a second invocation with a CUDA-event timer
(\texttt{\_capture.time\_kernel}, warmup 5, iters 50) for measured
perf.
\item Record one row per (kernel, ptx) with five static fields
(\texttt{register\_count}, \texttt{spill\_count},
\texttt{local\_memory\_bytes}, \texttt{instruction\_count},
\texttt{violations}) and two measured-perf fields
(\texttt{ms\_min}, \texttt{ms\_median}).
\end{enumerate}

\texttt{analysis/p4\_correlation.py} then pairs each correct kernel
with its buggy variant (by naming convention) and reports
$\Delta$regs, $\Delta$instrs, $\Delta$ms\_median, and $\Delta$perf\%.

\subsection{Assumptions}
\label{sec:method:assumptions}

The empirical claim depends on four assumptions.

\begin{enumerate}
\item Triton kernels are compiled fresh per run. Step 1 (cache clear)
removes any cross-run contamination of the PTX under test.
\item The representative shape per kernel is fixed for the perf timing.
Per-shape perf signatures are not measured here and are a noted
extension in Section~\ref{sec:limitations}.
\item The static metrics measured (registers, spills, instructions,
local memory) are an exhaustive enough proxy for the kernel's work
envelope at the granularity we test. We do not measure occupancy
directly.
\item The 9 paired correct and buggy variants represent both structural
and semantic LLM-style bugs. The companion paper~\cite{gpuemuP1}
establishes the bug taxonomy and the seeded transcription errors.
\end{enumerate}

\section{Evaluation}
\label{sec:eval}

\textbf{Setup.} vast.ai RTX~3060 for the single-GPU demonstration
below, image \texttt{pytorch/pytorch:2.4.0-cuda12.4-cudnn9-devel}.
The same driver and the same RTX~3060 run are then joined with four
additional GPU classes (A10, L40S, A100~SXM4, H100~NVL) for the
cross-architecture analysis in Section~\ref{sec:eval:cross}. The
canonical run identifier on RTX~3060 is
\texttt{run-20260611-142511-884321}; the other four are listed in
Table~\ref{tab:runs}.

\textbf{Static and perf per kernel} on RTX~3060 are reported in
Table~\ref{tab:perkernel} for a selected subset.

\begin{table}[h]
\centering
\caption{Per-kernel static metrics and measured perf on RTX~3060,
selected entries.}
\label{tab:perkernel}
\begin{tabular}{lrrrr}
\toprule
kernel & regs & spills & instrs & ms\_median \\
\midrule
gelu\_triton           & 232 & 0 & 205 & 0.099 \\
gelu\_triton\_buggy    & 224 & 0 & 197 & 0.102 \\
silu\_triton           & 136 & 0 & 109 & 0.097 \\
silu\_triton\_buggy    & 135 & 0 & 108 & 0.096 \\
rmsnorm\_triton        &  93 & 0 &  86 & 0.102 \\
rmsnorm\_triton\_buggy &  92 & 0 &  85 & 0.101 \\
softmax\_triton        & 130 & 0 & 116 & 0.103 \\
softmax\_triton\_buggy & 130 & 0 & 116 & 0.101 \\
\bottomrule
\end{tabular}
\end{table}

\textbf{Headline: paired diffs (buggy minus correct)}
are reported in Table~\ref{tab:paired}.

\begin{table}[h]
\centering
\caption{Paired diffs on RTX~3060, run \texttt{884321}. Static
$\Delta$ separates structural from semantic; measured
$\Delta$perf\% stays within $\pm 5\%$ for every pair.}
\label{tab:paired}
\resizebox{\textwidth}{!}{%
\begin{tabular}{lrrrrl}
\toprule
pair & $\Delta$regs & $\Delta$instrs & $\Delta$ms\_median & $\Delta$perf\% & bug class \\
\midrule
gelu\_triton $\to$ gelu\_triton\_buggy             & $\mathbf{-8}$ & $\mathbf{-8}$ & $+0.0031$ & $+3.2\%$ & structural (drops $0.5\times$ multiply) \\
matmul\_triton $\to$ matmul\_triton\_buggy         & $\mathbf{-15}$ & $\mathbf{-8}$ & $-0.0034$ & $-2.8\%$ & structural (\texttt{acc=} vs \texttt{acc+=}) \\
l2norm\_triton $\to$ l2norm\_triton\_buggy         & $-1$ & $-1$ & $+0.0004$ & $+0.3\%$ & structural (drops sqrt) \\
silu\_triton $\to$ silu\_triton\_buggy             & $-1$ & $-1$ & $-0.0005$ & $-0.6\%$ & micro-structural \\
rmsnorm\_triton $\to$ rmsnorm\_triton\_buggy       & $-1$ & $-1$ & $-0.0007$ & $-0.7\%$  & micro-structural \\
\textbf{leaky\_relu\_triton $\to$ leaky\_relu\_triton\_buggy} & $\mathbf{0}$ & $\mathbf{0}$ & $+0.0041$ & $+4.2\%$ & \textbf{semantic (constant only)} \\
\textbf{softmax\_triton $\to$ softmax\_triton\_buggy} & $\mathbf{0}$ & $\mathbf{0}$ & $-0.0023$ & $-2.2\%$ & \textbf{semantic (load-mask constant)} \\
\bottomrule
\end{tabular}%
}
\end{table}

\begin{figure}[h]
\centering
\includegraphics[width=\textwidth]{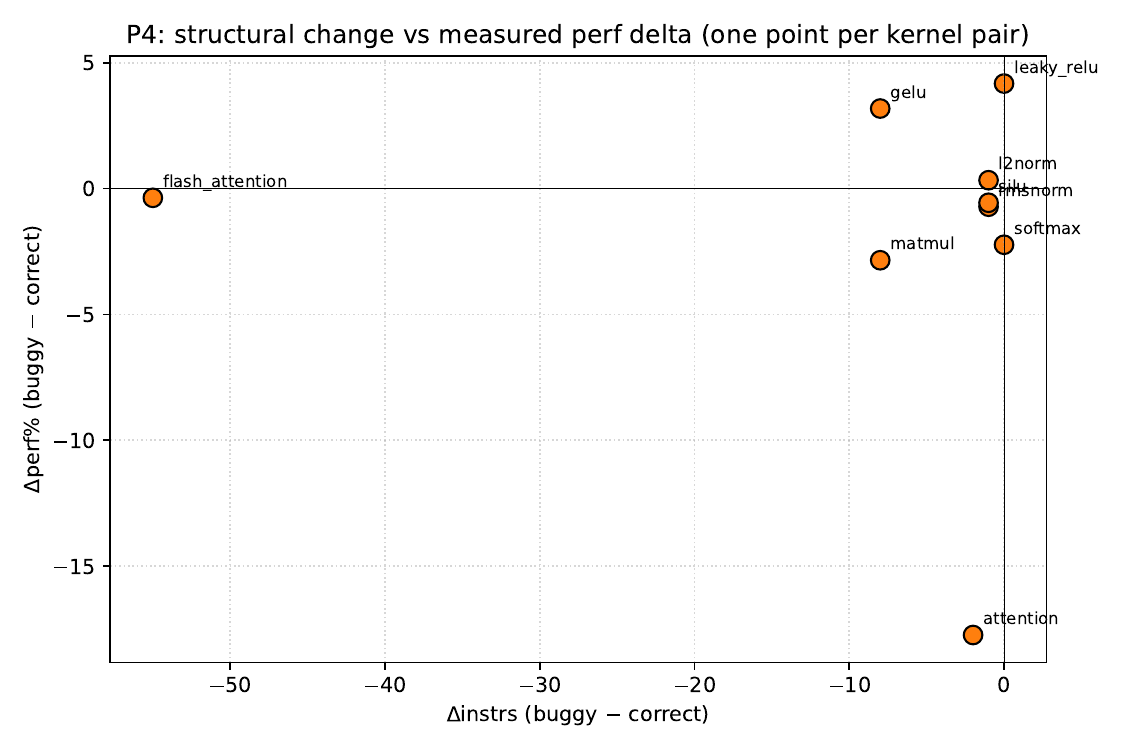}
\caption{$\Delta$instrs vs $\Delta$perf\% on RTX~3060, one point per
(correct, buggy) pair across all nine. The x-axis separates the
two bug classes cleanly; the y-axis does not.}
\label{fig:structperf}
\end{figure}

The pattern is clean on the static axis, noisy on the perf axis.

\textit{Structural bugs} (gelu missing $0.5$, matmul \texttt{acc=}
vs \texttt{acc+=}, l2norm and rmsnorm missing \texttt{sqrt}, silu
$\beta$ confusion) all show nonzero $\Delta$regs and $\Delta$instrs.
A simple static-delta CI gate ($|\Delta\text{regs}| \geq 1 \Rightarrow$
flag) catches all of them. Measured $\Delta$perf\% on the same pairs
ranges from $-2.8\%$ to $+3.2\%$, with no consistent sign: at the
sub-millisecond scale these kernels run, per-launch variance is the
dominant component.

\textit{Semantic-only bugs} (softmax \texttt{other=0.0} vs
\texttt{-inf}; leaky\_relu wrong $\alpha$ constant) compile to
byte-identical PTX (softmax: 130 registers, 116 instructions, byte-equal
\texttt{.ptx}). A static-delta gate is blind to them by construction.
Their measured $\Delta$perf\% ($-2.2\%$ and $+4.2\%$ respectively)
sits in the same $\pm 5\%$ noise band as the structural pairs, which
is why measured perf alone does not separate the two classes at this
kernel scale.

\subsection{Cross-architecture consistency}
\label{sec:eval:cross}

The single-GPU finding raises an obvious portability question. Does a
static-PTX gate calibrated on RTX~3060 transfer to data-center GPUs?
We re-ran the P4 driver on five GPU classes through the same vast.ai
harness: RTX~3060 (sm\_86), A10 (sm\_86), L40S (sm\_89),
A100~SXM4 (sm\_80), and H100~NVL (sm\_90). The headline result is
that static metrics are architecture-independent.

For each (correct, buggy) pair in this corpus, the Triton-emitted
PTX text our analyser parses produced identical static deltas across
all five GPU classes. The cross-GPU $\Delta$regs and $\Delta$instrs
table is therefore one column wide. This is a statement about the
PTX text metrics captured by the gpuemu daemon's artifact analyser,
not a general theorem about ptxas register allocation or SASS-level
resource usage across all targets. Table~\ref{tab:archindep} reports
it.

\begin{table}[h]
\centering
\caption{Cross-architecture static deltas. Identical across all five
GPU classes.}
\label{tab:archindep}
\resizebox{\textwidth}{!}{%
\begin{tabular}{lrrl}
\toprule
pair & $\Delta$regs & $\Delta$instrs & bug class \\
\midrule
gelu\_triton $\to$ gelu\_triton\_buggy                   & $\mathbf{-8}$  & $\mathbf{-8}$  & structural (drops $0.5\times$) \\
matmul\_triton $\to$ matmul\_triton\_buggy               & $\mathbf{-15}$ & $\mathbf{-8}$  & structural (\texttt{acc=} vs \texttt{acc+=}) \\
flash\_attention\_triton $\to$ flash\_attention\_triton\_buggy & $\mathbf{-36}$ & $\mathbf{-55}$ & structural (drops $\text{acc}\cdot\alpha$ rescale) \\
attention\_triton $\to$ attention\_triton\_buggy         & $\mathbf{-2}$  & $\mathbf{-2}$  & structural (drops $1/\sqrt{D}$ scale) \\
l2norm\_triton $\to$ l2norm\_triton\_buggy               & $-1$ & $-1$ & structural (drops sqrt) \\
rmsnorm\_triton $\to$ rmsnorm\_triton\_buggy             & $-1$ & $-1$ & structural (drops sqrt) \\
silu\_triton $\to$ silu\_triton\_buggy                   & $-1$ & $-1$ & micro-structural \\
\textbf{leaky\_relu\_triton $\to$ leaky\_relu\_triton\_buggy} & $\mathbf{0}$ & $\mathbf{0}$ & \textbf{semantic (constant only)} \\
\textbf{softmax\_triton $\to$ softmax\_triton\_buggy}    & $\mathbf{0}$ & $\mathbf{0}$ & \textbf{semantic (load-mask constant)} \\
\bottomrule
\end{tabular}%
}
\end{table}

The static-gate verdict transfers cleanly. Anything that flags or
passes on RTX~3060 flags or passes the same way on every other GPU
class. The two semantic-bug pairs (\texttt{leaky\_relu\_triton\_buggy},
\texttt{softmax\_triton\_buggy}) compile to identical PTX on all five
GPUs and are therefore invisible to any static gate everywhere. This is
the architecture-independent restatement of the main claim.

\begin{figure}[h]
\centering
\includegraphics[width=\textwidth]{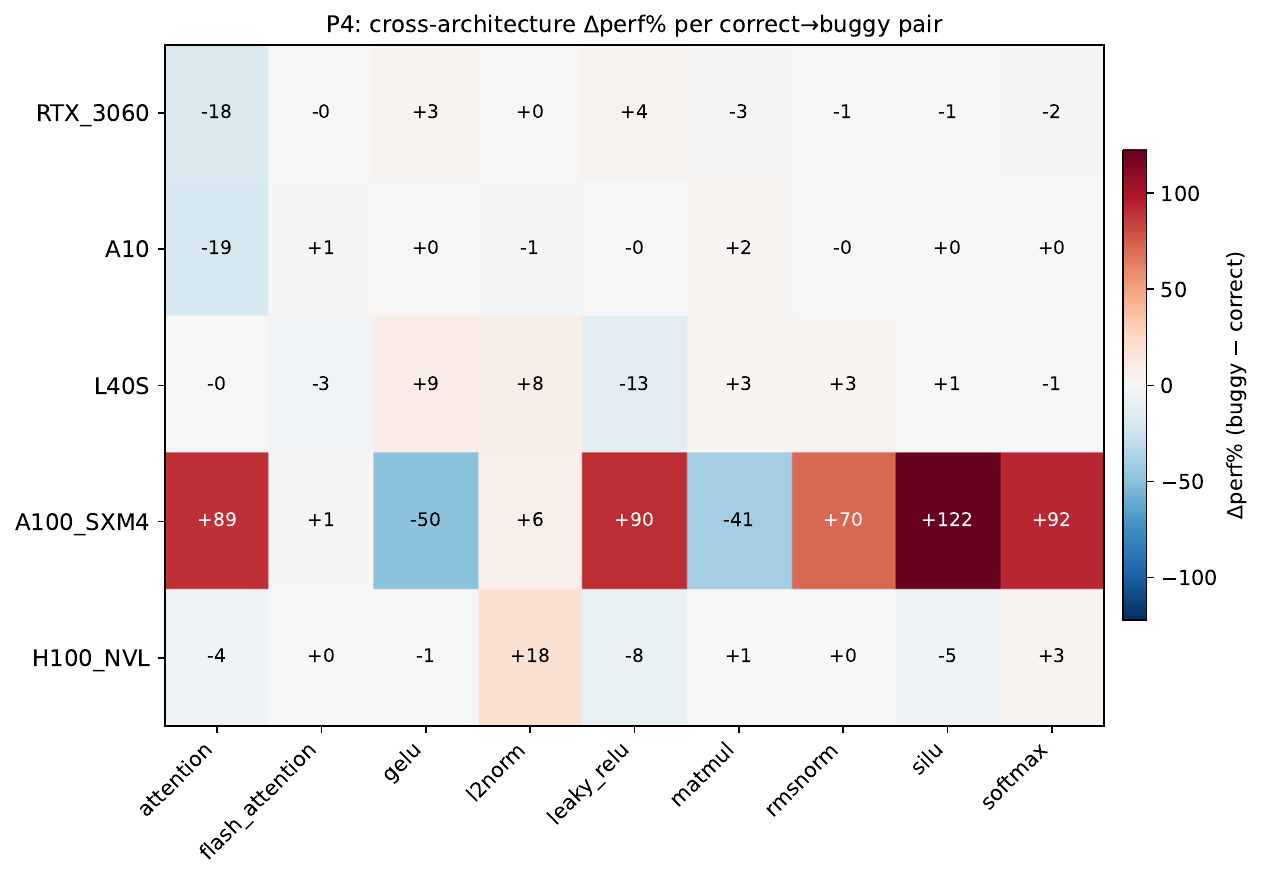}
\caption{Cross-architecture $\Delta$perf\% per pair. Only the
measured signal varies; static $\Delta$ is identical across all
five GPUs.}
\label{fig:crossgpu}
\end{figure}

\textbf{Measured $\Delta$perf\% varies across architectures and is
dominated by launch-overhead noise at the sub-millisecond scale.} For
completeness Table~\ref{tab:perfgrid} reports the per-architecture
$\Delta$perf\%. Figure~\ref{fig:crossgpu} plots the same data on a
divergent colormap so the A100 row's variance is visible.

\begin{table}[h]
\centering
\caption{Cross-architecture $\Delta$perf\% per pair. Four
architectures stay within $\pm 20\%$ of zero; A100~SXM4 is
dominated by shared-host launch variance.}
\label{tab:perfgrid}
\begin{tabular}{lrrrrr}
\toprule
pair & RTX~3060 & A10 & L40S & A100~SXM4 & H100~NVL \\
\midrule
attention\_triton $\to$ \_buggy           & $-17.7\%$ & $-19.2\%$ & $-0.0\%$  & $+89.0\%$ & $-3.9\%$ \\
flash\_attention\_triton $\to$ \_buggy    & $-0.4\%$  & $+1.1\%$  & $-3.0\%$  & $+1.0\%$  & $+0.3\%$ \\
gelu\_triton $\to$ \_buggy                & $+3.2\%$  & $+0.2\%$  & $+9.4\%$  & $-50.0\%$ & $-0.8\%$ \\
l2norm\_triton $\to$ \_buggy              & $+0.3\%$  & $-1.0\%$  & $+8.1\%$  & $+6.0\%$  & $+18.4\%$ \\
leaky\_relu\_triton $\to$ \_buggy         & $+4.2\%$  & $-0.2\%$  & $-12.5\%$ & $+89.7\%$ & $-8.1\%$ \\
matmul\_triton $\to$ \_buggy              & $-2.8\%$  & $+2.2\%$  & $+3.1\%$  & $-41.2\%$ & $+0.6\%$ \\
rmsnorm\_triton $\to$ \_buggy             & $-0.7\%$  & $-0.1\%$  & $+2.7\%$  & $+70.5\%$ & $+0.4\%$ \\
silu\_triton $\to$ \_buggy                & $-0.6\%$  & $+0.3\%$  & $+0.5\%$  & $+122.2\%$& $-5.1\%$ \\
softmax\_triton $\to$ \_buggy             & $-2.2\%$  & $+0.2\%$  & $-0.9\%$  & $+92.3\%$ & $+3.1\%$ \\
\bottomrule
\end{tabular}
\end{table}

The operational takeaway is that static-PTX gating is the portable
signal. $\Delta$regs and $\Delta$instrs are architecture-independent,
decided at compile time, and indistinguishable across the five GPU
classes for the same kernel. Measured $\Delta$perf\% is not portable at
the sub-millisecond scale these kernels occupy, so a perf-based gate
calibrated on one GPU class cannot be assumed to threshold meaningfully
on another without per-architecture re-calibration.

The five run records are listed in Table~\ref{tab:runs}. Each row
is independently replayable from the public gpuemu repository with
\texttt{python3 scripts/replay\_from\_b2.py --run-id <id>}, which
fetches \texttt{results.jsonl} and \texttt{summary.json} from the
B2 bucket and re-runs the analysis scripts without requiring GPU
access. The Backblaze path for each row is
\texttt{gpuemu/<run-id>/\{results.jsonl,summary.json\}}.

\begin{table}[h]
\centering
\caption{Run records on B2 bucket
\texttt{sarkar-dipankar-research}, prefix \texttt{gpuemu/}.}
\label{tab:runs}
\begin{tabular}{ll}
\toprule
GPU class & run id \\
\midrule
RTX 3060   & \texttt{run-20260611-142511-884321} \\
A10        & \texttt{run-20260611-145820-13358e} \\
L40S       & \texttt{run-20260611-144339-6247e5} \\
A100 SXM4  & \texttt{run-20260611-144659-0378af} \\
H100 NVL   & \texttt{run-20260611-145211-2c6308} \\
\bottomrule
\end{tabular}
\end{table}

\section{Discussion}
\label{sec:discussion}

This is a bounded claim with concrete implications for CI design.

A static-PTX-delta gate is a cheap useful pre-filter. It catches the
structural bugs at zero hardware cost. It is useful at PR time before
any GPU is provisioned.

It must not be the sole correctness gate. Semantic-only bugs slip
through. These bugs are often the most pernicious because they
preserve performance metrics. The fuzzing oracle from the companion
paper~\cite{gpuemuP1} remains necessary.

The two methodologies are complementary, not competing. A strong CI
pipeline runs both, with static gating as the fast filter and fuzzing
as the deep check.

The observed pattern refines the long-standing folk model. Static
instruction and register deltas are a useful indicator that the
compiled kernel changed structurally, but they are not a reliable
quantitative predictor of runtime at this kernel scale. Structural
changes may be performance-relevant, but the measured effect can be
masked by launch overhead and shared-host variance. Constant-only
semantic bugs are worse for CI: they preserve both static metrics
and performance-like signals, so they require an independent
correctness oracle.

\section{Limitations}
\label{sec:limitations}

The corpus has 9 paired correct and buggy variants. This is adequate
for the dichotomy demonstration and for the cross-architecture
portability check in Section~\ref{sec:eval:cross}, but insufficient
for fitting a quantitative perf-prediction model.

The static-metric portability finding is exact: the same PTX implies
the same \texttt{register\_count} and the same \texttt{instruction\_count}
regardless of GPU. The measured-$\Delta$perf\% portability finding is
the opposite. At the sub-millisecond scale these corpus kernels occupy,
per-launch variance (particularly on shared-host SXM4 inventory)
dominates the few-instruction differences. Strengthening $\Delta$perf\%
as a portable signal would require either longer-running kernels (larger
shapes) or per-architecture re-calibration.

Perf timing uses a single representative shape per kernel. Perf varies
by shape. A richer P4 sweep would parameterise over shape and produce
a per-shape perf signature on each GPU class.

The default \texttt{ArtifactCheckConfig} threshold
(\texttt{max\_registers = 64}) flags every modern Triton kernel as
\texttt{ExcessiveRegisters}. This is a configuration concern, not a
research finding: the threshold is a left-over default that should
be retuned per target before the static-PTX gate is enabled in CI.

\section{Conclusion}
\label{sec:conclusion}

Static PTX deltas track structural source changes in this corpus
and toolchain. They are useful as a portable pre-filter for CI:
kernels that skip work register a nonzero $\Delta$regs or
$\Delta$instrs, and kernels that swap one constant for another
compile to identical PTX. Measured runtime deltas at the
sub-millisecond scale these corpus kernels occupy remain hardware-
and noise-sensitive: the same structural change registered $-50\%$
on A100~SXM4 and $+9.4\%$ on L40S in our cross-architecture sweep
(Table~\ref{tab:perfgrid}). The cross-architecture sweep in
Section~\ref{sec:eval:cross} therefore strengthens only the static
side. The static signal ($\Delta$regs, $\Delta$instrs) for each pair
in this corpus is identical across the five GPU classes we measured.
A perf-based gate needs per-architecture calibration before it is
used as anything more than a smoke check. The data supports an
integrated CI design that runs both: cheap portable static checks at
PR time, and fuzzing-based correctness
checks~\cite{gpuemuP1} before merge.

\paragraph{Artefact.}
{\sloppy The corpus, the static-plus-perf driver
(\texttt{p4\_artifacts.py}), the correlation analysis
(\texttt{p4\_correlation.py}), and the
\texttt{replay\_from\_b2.py} script that fetches each cited run
record are bundled in the public \textsf{gpuemu-corpus} package at
\url{https://github.com/sarkar-dipankar/gpuemu-corpus}. The
validator daemon and the artefact analyser are at
\url{https://github.com/Skelf-Research/gpuemu}.\par}

\paragraph{License.}
This preprint is released under
\href{https://creativecommons.org/licenses/by/4.0/}{CC-BY 4.0}.

\bibliographystyle{splncs04}
\bibliography{refs}

\end{document}